\documentclass[preprint,prd,aps,showpacs,showkeys,nofootinbib]{revtex4-1}
\usepackage{amssymb}
\usepackage{amsmath}
\usepackage{graphicx}
\usepackage{subfigure}
\usepackage{float}
\usepackage{mathrsfs}
\usepackage{amsmath}
\usepackage{xcolor}
\newcommand{\be}{\begin{eqnarray}}
\newcommand{\ee}{\end{eqnarray}}

\begin{document}


\title{The LISA-Taiji network: precision localization of massive black hole binaries}

\author{Wen-Hong Ruan$^{1,2}$}
\email{ruanwenhong@itp.ac.cn}

\author{Chang Liu$^{1,2}$}
\email{liuchang@itp.ac.cn}

\author{Zong-Kuan Guo$^{1,2}$}
\email{guozk@itp.ac.cn}

\author{Yue-Liang Wu$^{1,2,3}$}
\email{ylwu@itp.ac.cn}

\author{Rong-Gen Cai$^{1,2}$}
\email{cairg@itp.ac.cn}

\affiliation{$^1$CAS Key Laboratory of Theoretical Physics, Institute of Theoretical Physics,
 Chinese Academy of Sciences, P.O. Box 2735, Beijing 100190, China}
\affiliation{$^2$School of Physical Sciences, University of Chinese Academy of Sciences,
 No.19A Yuquan Road, Beijing 100049, China}
\affiliation{$^3$International Centre for Theoretical Physics Asia-Pacific, Beijing, China}

\begin{abstract}
A space-based gravitational-wave detector, LISA, consists of a triangle of three spacecrafts
with a separation distance of 2.5 million kilometers in a heliocentric orbit behind the Earth.
Like LISA, Taiji is compose of a triangle of three spacecrafts
with a separation distance of 3 million kilometers in a heliocentric orbit ahead of the Earth.
They are expected to launch in 2030-2035.
Assuming a one-year overlap,
we propose the LISA-Taiji network in space
to fast and accurately localize the gravitational-wave sources.
We use the Fisher information matrix approach to analyze the sky localization for coalescing massive black hole binaries.
For an equal-mass black hole binary located at redshift of 1 with a total intrinsic mass of $10^5 M_{\odot}$,
the LISA-Taiji network may achieves about four orders of magnitude improvement on the event localization region
compared to an individual detector.
The precision measurement of sky location from the gravitational-wave signal may completely identify the host galaxy with low redshifts
prior to the final black hole merger.
Such the identification of the host galaxy is helpful for the follow-up change in electromagnetic emissions of the accretion disk
when the massive black hole binary merges to a single massive black hole,
and enables the coalescing massive black hole binaries to be used as a standard siren.
\end{abstract}

\maketitle

\begin{figure}[H]
\center
\includegraphics[height=7cm]{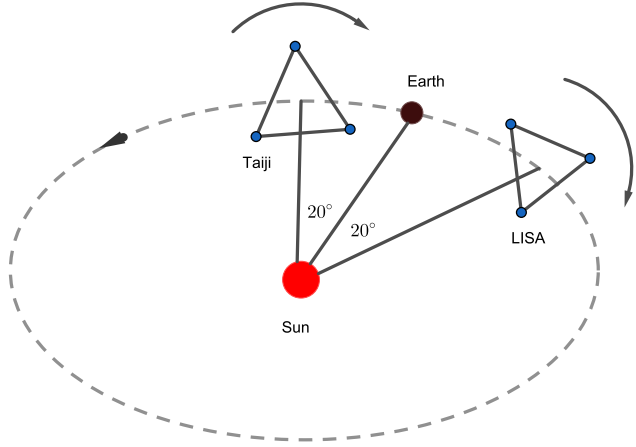}
\caption{Configuration of the LISA-Taiji network.
LISA consists a triangle of three spacecrafts with a separation distance of 2.5 million kilometers
in a heliocentric orbit behind the Earth by about $20^{\circ}$
while Taiji has a 3 million kilometers separation in a heliocentric orbit ahead of the Earth by about $20^{\circ}$.
The LISA-Taiji network with a separation distance of about $0.7$ AU,
can improve the sky localization of coalescing black hole binaries.}
\label{fig:configuration}
\end{figure}

LISA, a space-based gravitational wave (GW) observatory,
was proposed in 1990s to detect GWs with a frequency band from $10^{-4}$ Hz to $10^{-1}$ Hz.
LISA consists of a triangle of three spacecrafts with a separation distance of 2.5 million kilometers
in orbit around the Sun,
which bounce lasers between each other.
The constellation fellows the Earth by about $20^{\circ}$ (Fig.~\ref{fig:configuration}).
It is expected to launch in 2030-2035, with a mission lifetime of 4 years extendable to 10 years~\cite{Audley:2017drz}.
Like LISA, Taiji is compose of a triangle of three spacecrafts with 3 million kilometers separations
in a heliocentric orbit ahead of the Earth by about $20^{\circ}$ (Fig.~\ref{fig:configuration}).
Compared to LISA, Taiji is sensitive to low-frequency GWs (see Methods).
Taiji would launch during the same period~\cite{Hu:2017mde}.
Assuming a one-year overlap, if Taiji joins the LISA constellation,
the LISA-Taiji network in space (Fig.~\ref{fig:configuration})
can significantly improve the sky localization of GW sources, including the luminosity distance and solid angle,
due to the faraway seperation of the two constellations.

For a transient GW signal from a stellar-mass black hole binary,
it is hard to determine the sky location of the GW source using a single ground-based GW detector
because detectors are sensitive to GWs from nearly all directions.
With two detectors at distinct locations,
the position of the source can in principle be restricted to an annulus in the sky
by triangulation using the time difference on arrival at the two detectors.
A network of more than two detectors can localize the sky position of the source
using the arrival time difference with the help of the phase difference and amplitude ratios of GWs on arrival at the detectors.
For example, the sky localization of GW170814 is significantly improved due to the joining of the Advanced Virgo detector,
reducing the area of the 90\% credible region from 1160 deg$^2$ using only the two
Advanced LIGO detectors to 60 deg$^2$ using the LIGO-Virgo network~\cite{Abbott:2017oio}.
The joining of the Advanced Virgo detector has played an important role in sky localization.

For GWs from coalescing massive black hole binaries (MBHBs) with total masses between $10^4 M_{\odot}$ and $10^8 M_{\odot}$,
which are expected to be the strongest GW sources for space-based GW observatories,
a single detector can localize the sky position of the MBHB by the motion of the detector in space.
If the host galaxy is identified from GW observations,
one can easily have the redshift of the source with a good accuracy by optical measurements.
By measuring these GWs as potentially powerful standard sirens~\cite{Schutz:1986gp,Abbott:2017xzu,Chen:2017rfc},
we will have detailed information on the high-redshift expansion history of the Universe.

Actually there are indirect evidences for the existence of MBHBs in galactic centers.
Although the origin of these massive black holes powering active galactic nuclei is unknown,
MBHBs inevitably form due to frequent galaxy mergers~\cite{Begelman:1980vb}.
MBHBs with kpc scale separations have been unambiguously detected
in optical and X-ray surveys~\cite{Komossa:2002tn}.
However, observations of MBHBs with sub-pc scale separations is particularly challenging
because these small separations at cosmic distance are well below the angular resolving power of the current telescopes.
In this case, only MBHB candidates have been found through optical variability with the periods of $\sim 14$ years in the center of Ark 120~\cite{Li:2016hcm}
and $\sim 20$ years in the center of NGC 5548~\cite{Li:2017eqf}.
Fortunately, when the orbital period of the system becomes smaller than hours,
there is a good chance to detect MBHBs in galactic centers by GW measurements.
Furthermore, space-based GW detectors would observe GWs generated by the coalescence with a high signal-to-noise ratio.
If the host galaxy is identified by GW detections prior to the final black hole merger,
it is helpful for the follow-up change in electromagnetic emissions of the accretion disk
when the MBHB merges to a single black hole,
and enables the coalescing MBHB to be used as a standard siren.

The coalescence of MBHBs with total masses from $10^4 M_{\odot}$ to $10^8 M_{\odot}$
in general lasts for several days, months or even years in the frequency band of LISA and Taiji.
Due to the motion of the detector in space,
the time dependence of the antenna pattern functions
plays a crucial role in localizing the position of the GW source.
Hence, a single space-based detector can be effectively treated as a network
including a set of detectors at different locations along the detector's trajectory in space,
which observe a given GW event at different time.
LISA is expected to localize GW sources to the angular resolution of $1 - 100$ deg$^2$,
which depends on the mass, distance and inclination angle of GW sources~\cite{Cutler:1997ta,Audley:2017drz}.
Such the angular resolution is not good enough to identify the source galaxy.
If Taiji joins LISA,
the LISA-Taiji network can significantly improves the sky localization of GW sources
by triangulation using the time difference on arrival at the two detectors (see Methods).

\begin{figure}[H]
\center
\includegraphics[height=7cm]{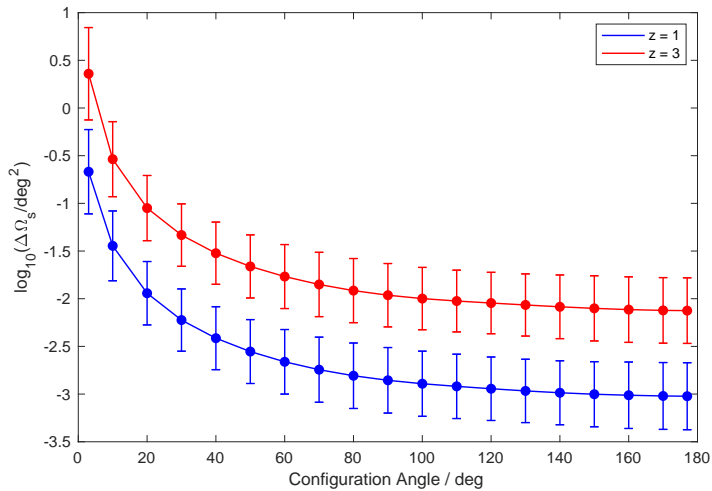}
\caption{Measurements of the angular resolution, $\Delta \Omega_s$,
as a function of the configuration angle subtended by the heliocentric orbit between two detectors in the LISA-Taiji network.
We choose an equal-mass black hole binary with a total intrinsic mass of $10^5 M_{\odot}$, located at redshifts of $z = 1$ (blue) and $z = 3$ (red).
The $1\sigma$ uncertainties are evaluated using a catalogue of 10,000 simulated sources at different sky positions (see Methods).}
\label{fig:angle}
\end{figure}

Using the Fisher information matrix approach (see Methods),
we analyze the sky localization for coalescing MBHBs in the LISA-Taiji network.
From Fig.~\ref{fig:angle} we can see that the angular resolution is a function of the configuration angle, $\beta$,
subtended by the heliocentric orbit between LISA and Taiji.
When the configuration angle becomes $180^{\circ}$,
as expected the angular resolution reaches a minimum value.
For an equal-mass black hole binary with a total intrinsic mass of $10^5 M_{\odot}$, located at redshifts of $z = 1$ and $z = 3$,
the angular resolution is improved by about 2 orders of magnitude as the configuration angle varies from $0^{\circ}$ to $40^{\circ}$
while it is improved by about 0.6 order of magnitude from $\beta=40^{\circ}$ to $\beta=180^{\circ}$.
Hence, the LISA-Taiji network with $\beta=40^{\circ}$ can effectively help us to fast and accurately localize the GW sources.
In what follows, our analysis is based on $\beta=40^{\circ}$.

\begin{figure}[H]
\center
\includegraphics[height=5.3cm]{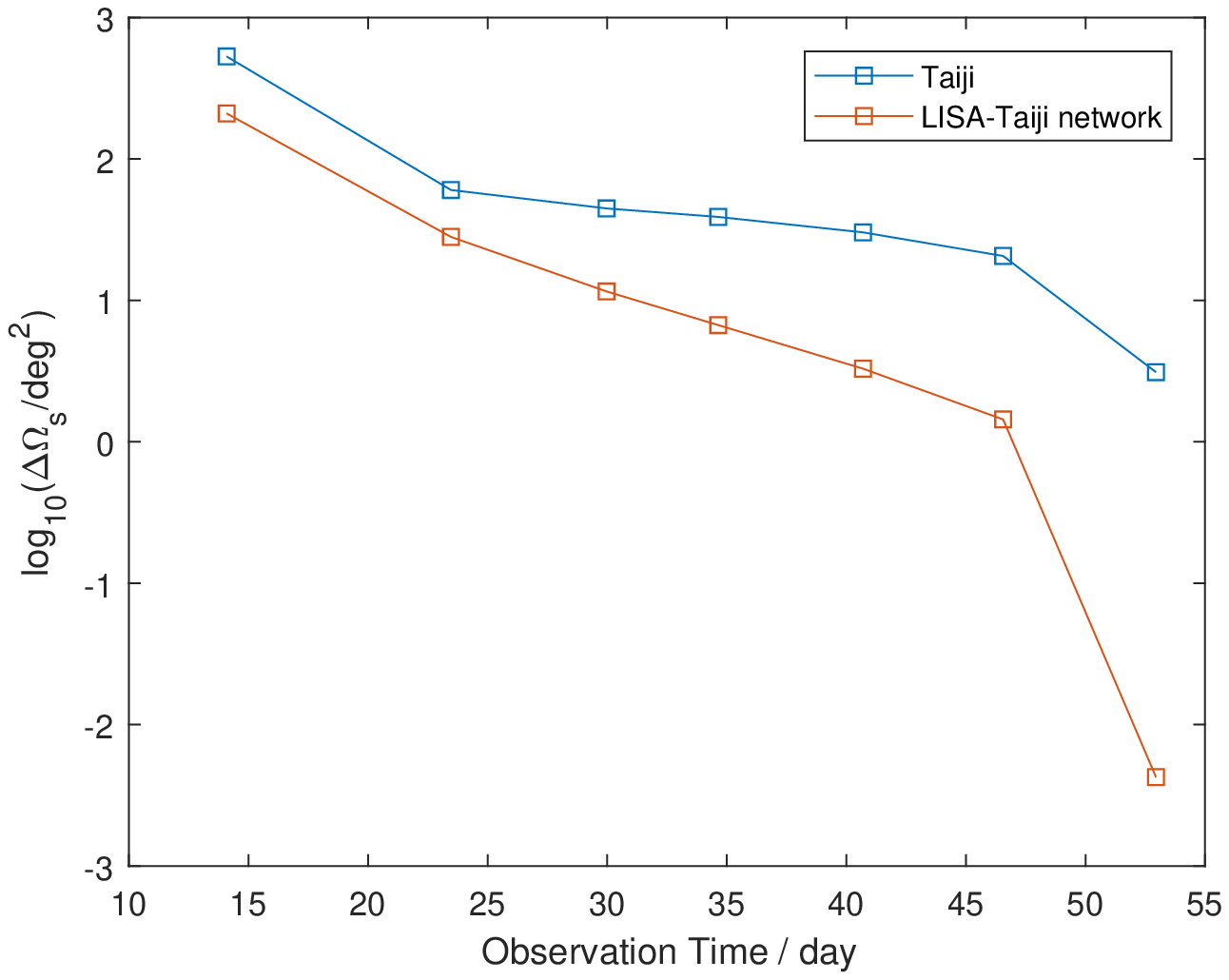}
\includegraphics[height=5.3cm]{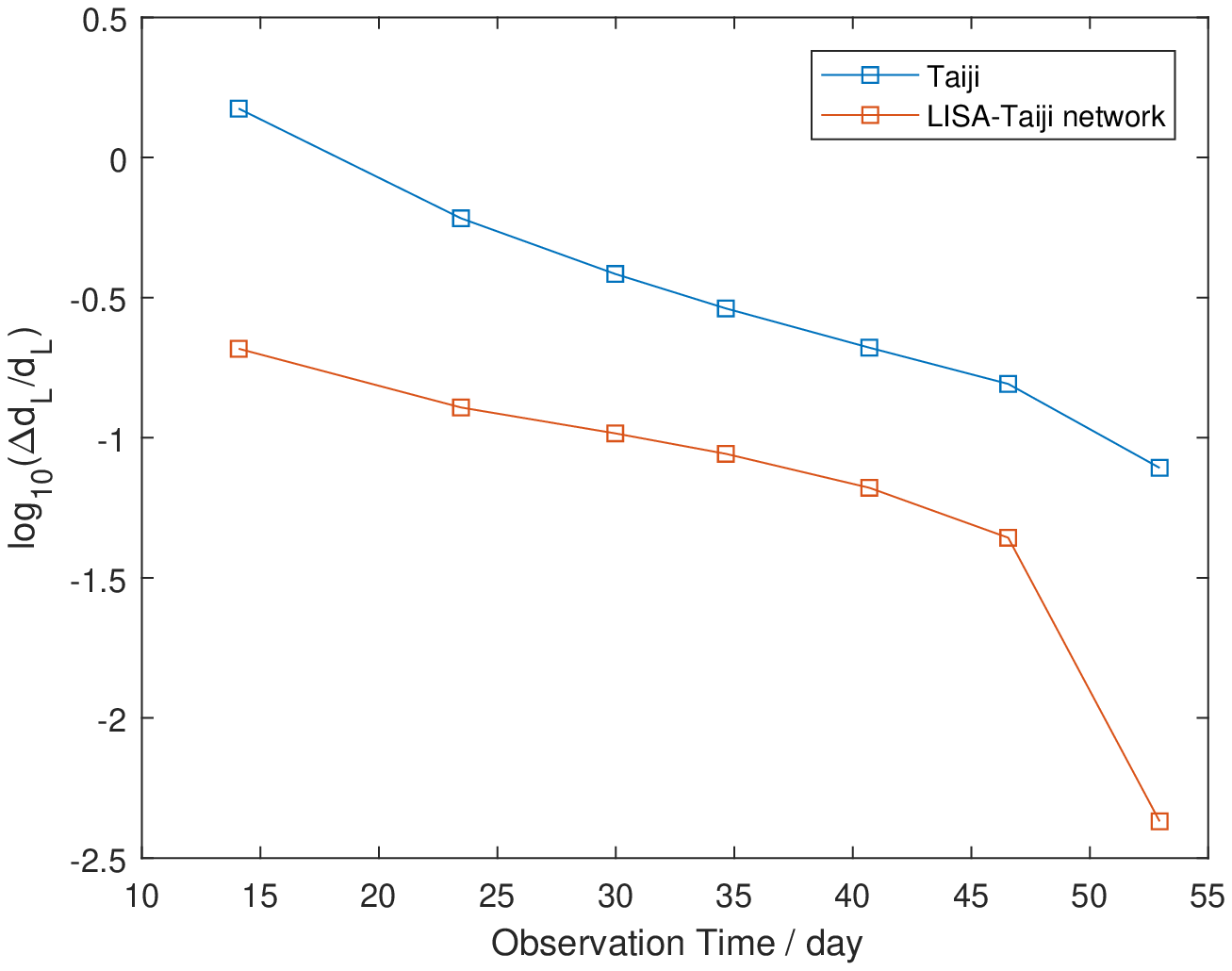}
\caption{Measurements of the angular resolution, $\Delta \Omega_s$, (left panel) and the luminosity distance uncertainty, $\Delta d_L/d_L$, (right panel)
as a function of observation time in Taiji (blue) and the LISA-Taiji network (red).
We choose an equal-mass black hole binary, located at redshift of $z = 1$ with a total intrinsic mass of $10^5 M_{\odot}$.}
\label{fig:time}
\end{figure}

We consider an equal-mass black hole binary, located at redshift of $z = 1$ with a total intrinsic mass of $10^5 M_{\odot}$.
In Fig.~\ref{fig:time}, we show measurements of the angular resolution (left panel) and the luminosity distance uncertainty (right panel)
as a function of observation time in Taiji (blue) and the LISA-Taiji network (red).
In Taiji the source can be localized with $\Delta \Omega_s < 4$ deg$^2$
and $\Delta d_L/d_L < 8 \%$,
while in the LISA-Taiji network the source can be localized with $\Delta \Omega_s < 0.005$ deg$^2$
and $\Delta d_L/d_L < 0.5 \%$.
The constraints on the solid angle are improved by three orders of magnitude
and the luminosity distance are improved by one order of magnitude.
Therefore, the LISA-Taiji network may achieves about four orders of magnitude improvement on the source localization region
compared to an individual detector.

\begin{figure}[H]
\center
\includegraphics[height=5.3cm]{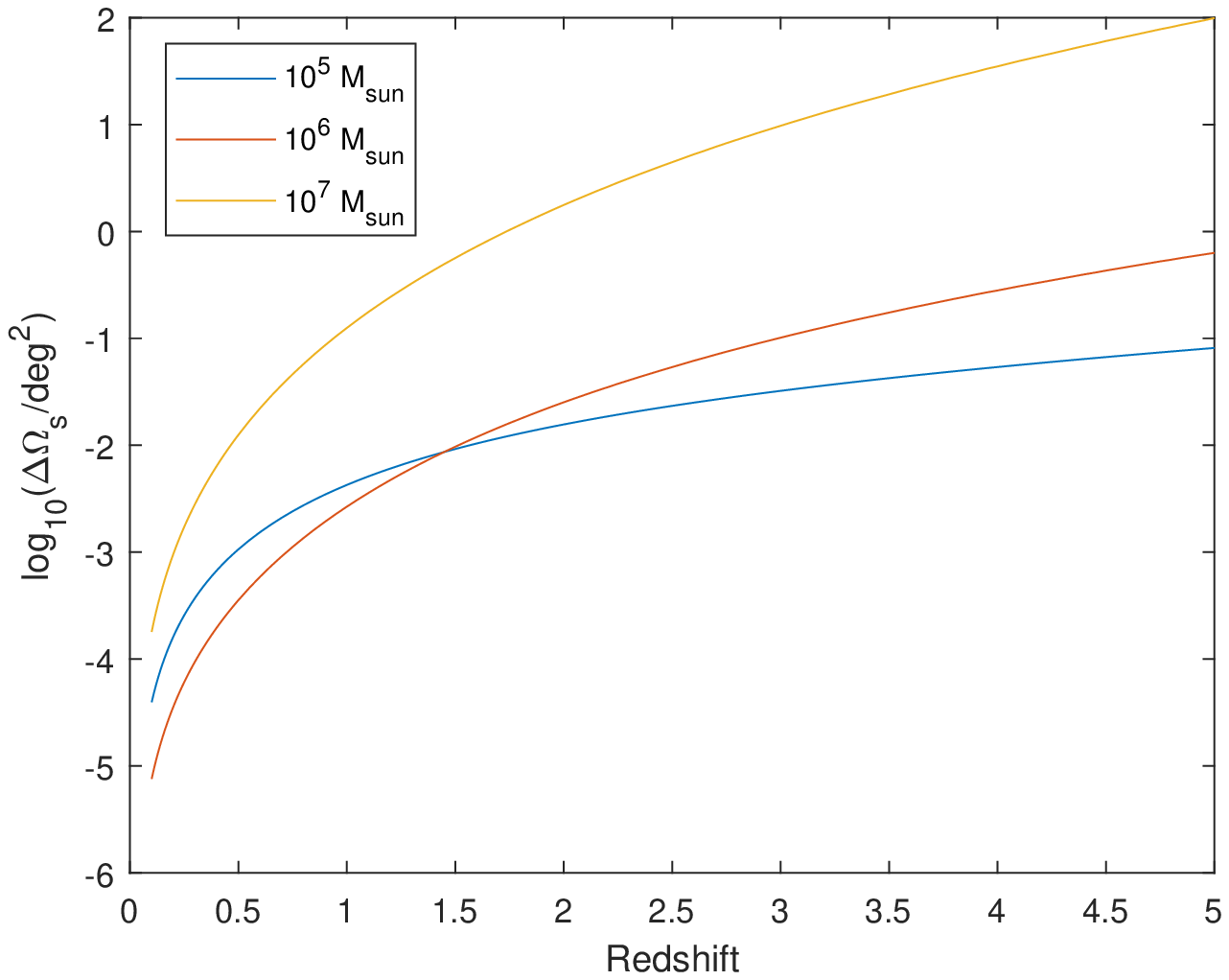}
\includegraphics[height=5.3cm]{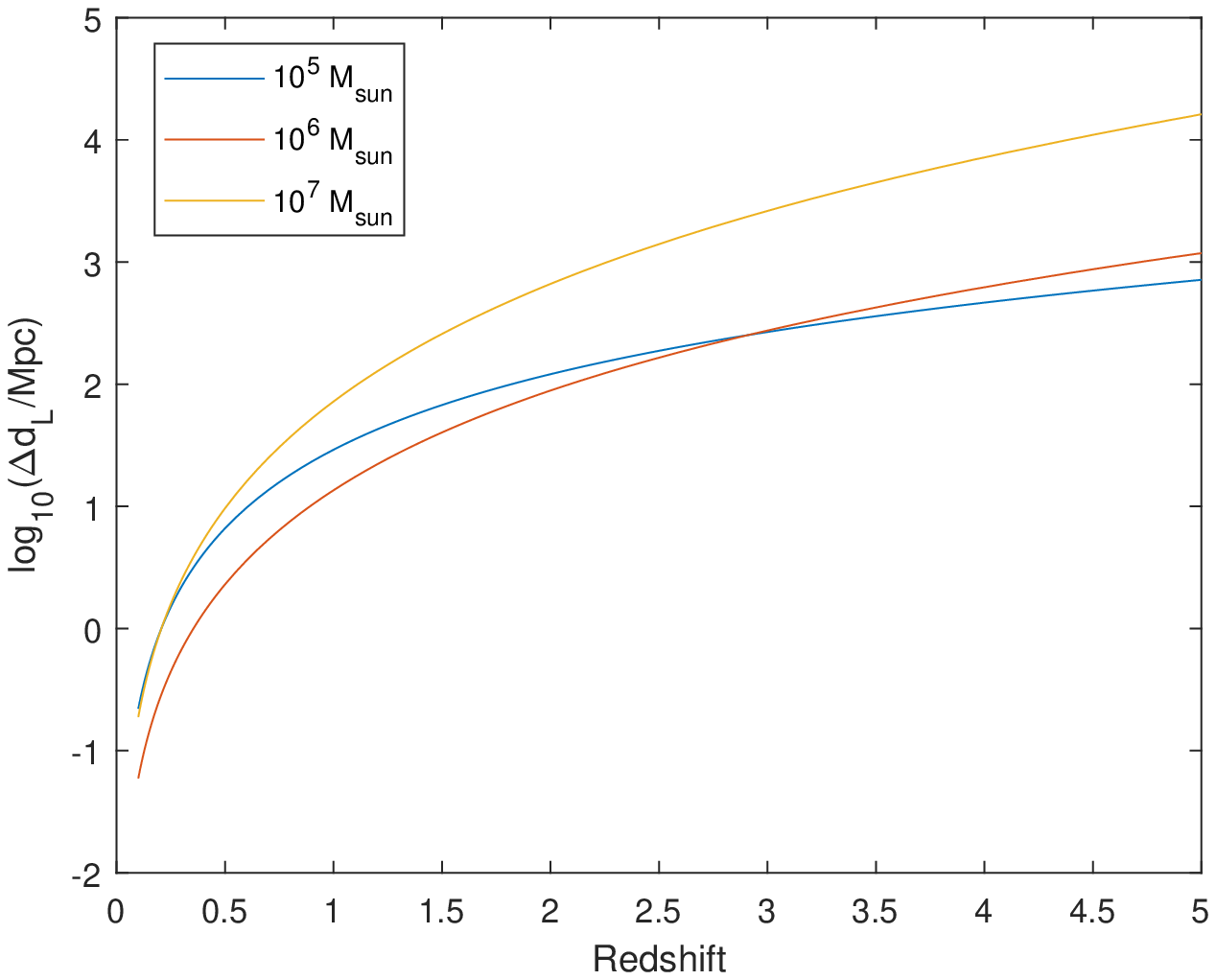}
\caption{Measurements of the angular resolution (left panel) and the luminosity distance uncertainty (right panel),
as a function of redshifts of the equal-mass black hole binaries
with total intrinsic masses of $10^5 M_{\odot}$ (blue), $10^6 M_{\odot}$ (red) and $10^7 M_{\odot}$ (yellow).}
\label{fig:redshift}
\end{figure}

We find that the precision measurement of sky location in the LISA-Taiji network may completely identify the host galaxy prior to the final black hole merger.
The unique identification of the host galaxy allows us to measure the evolution history of the Universe via the distance-redshift relation.
In Fig.~\ref{fig:redshift} we show measurements of the angular resolution (left panel) and the luminosity distance uncertainty (right panel),
as a function of redshifts of the equal-mass black hole binaries
with total intrinsic masses of $10^5 M_{\odot}$ (blue), $10^6 M_{\odot}$ (red) and $10^7 M_{\odot}$ (yellow).
Assuming that galaxies are uniformly distributed in comoving volume with a number density of $0.02$ Mpc$^{-3}$,
we estimate the number of potential galaxies within the even localization volume.
We find that the LISA-Taiji network can identify the host galaxy of the MBHB with a total intrinsic mass of $10^5 M_{\odot}$
if the galaxy redshift is smaller than 0.75, with a total intrinsic mass of $10^6 M_{\odot}$ if the galaxy redshift is smaller than 0.96 and with a total intrinsic mass of $10^7 M_{\odot}$ if the galaxy redshift is smaller than 0.45.

We have investigated the ability of the LISA-Taiji network to localize the GW sources of MBHBs
using the Fisher information matrix method.
We find the LISA-Taiji network achieves a remarkable ability improvement on the sky localization compared to an individual detector.
It is possible to identify host galaxies only from GW detections prior to merger.
It provides us a good chance to measure possible change of electromagnetic emissions of the accretion disk when the MBHB merges to a single massive black hole,
and allows us to explore the expansion of the Universe using MBHBs as standard sirens even without consequently electromagnetic variability.

\noindent{\bf Acknowledgments} We acknowledge helpful conversations with Wen Zhao, Jian-Min Wang and Jan Zaanen.
ZKG is supported in part by the National Natural Science Foundation of China Grants
No.11690021 and No.11575272,
in part by the Strategic Priority Research Program of the Chinese Academy of Sciences Grant No. XDB23030100,
No. XDA15020701 and by Key Research Program of Frontier Sciences, CAS.
YLW is supported by the National Natural Science Foundation of China Grants
No.11851302 and No.11747601.
RGC is supported by the National Natural Science Foundation of China Grants
No.11690022, No.11435006 and No.11821505.

\vspace{5mm}
\noindent \textcolor[rgb]{0,0,0.6}{\bf METHODS}\\
\noindent{\bf GW waveforms and detector response functions.}
The matched filter is used to search for the GW signal from data and estimate the parameters of the GW source,
which requires the waveform template of coalescing compact binaries.
The GW signal from an inspiraling nonspinning MBHB can be modeled by a restricted post-Newtonian (PN) waveform
(i.e., the amplitude is kept at the dominant Newtonian level while the phase is evolved to the second PN order).
Two polarization amplitudes of the GW signal are given by
\be
\label{1}
  h_{+,\times}(t) &=& \frac{2G M_c \eta^{2/5}x}{c^2 d_L} \Big [ H_{+,\times}^{(0)}+x^{1/2}H_{+,\times}^{(1/2)}
  +xH_{+,\times}^{(1)}+x^{3/2}H_{+,\times}^{(3/2)}+x^{2}H_{+,\times}^{(2)} \nonumber \\
  && +x^{5/2}H_{+,\times}^{(5/2)} + {\cal O} \left(\frac{1}{c^6}\right)\Big],
\ee
where $d_L$ is the luminosity distance to the source,
$M_c = \eta^{3/5} M$ is the chirp mass,
$M=m_1+m_2$ is the total mass
and $\eta=m_1 m_2/M^2$ is the symmetric mass ratio.
The invariant PN velocity parameter $x$ is defined by
\begin{equation}\label{2}
  x\equiv\left(\frac{GM\omega}{c^3}\right)^{2/3},
\end{equation}
where $\omega$ is the orbital frequency of the binary for a circular orbit.
To the lowest PN order in the amplitude evolution, the waveform is given for $t<t_c$ by
\be
\label{3}
  h_{+}(t) &=& -\frac{1+{\rm cos}^2\iota}{2}\left(\frac{G M_c}{c^2 d_L}\right)\left(\frac{t_c-t}{5G M_c /c^3}\right)^{-1/4}
   \cos [2\phi_c+2\phi(t-t_c;M_c,\eta)], \\
  h_{\times}(t) &=& -{\rm cos}\iota\left(\frac{G M_c}{c^2 d_L}\right)\left(\frac{t_c-t}{5G M_c /c^3}\right)^{-1/4}
   {\rm sin}[2\phi_c+2\phi(t-t_c;M_c,\eta)],
\label{4}
\ee
where $\iota$ is the angle between the orbital angular momentum axis of the binary and the direction to the detector,
$t_c$ and $\phi_c$ is the coalescence time and coalescence phase.
In the LISA-Taiji network, we choose the polar coordinate system with the sun as its origin.
So the strain on a detector is given by
\be
h(t+\tau) = F_{+}(\theta,\phi,\iota,\psi;t)h_{+}(t)
 +F_{\times}(\theta,\phi,\iota,\psi;t)h_{\times}(t),
\ee
where $F_{+}$ and $F_{\times}$ are the detector response functions,
$\theta$ and $\phi$ are the colatitude and longitude of the binary in the polar coordinate system
(assuming that the center-of-mass of the binary is at rest),
and $\psi$ is the polarization angle.
Here, $\tau$ is the delay between the arrival time of GWs at the Sun and the arrival time at the detector,
which is given by
\begin{equation}
\tau=\frac{{\vec{x}(t)\cdot \hat{k}}}{c},
\end{equation}
where ${\vec x}(t)$ is the position vector of the source relative to the detector and
$\hat k$ is the unit vector from the source to the Sun.
Therefore, the strain can be written as
\begin{equation}
\label{eq:strain}
h(t)=-\frac{GM_c}{c^2 D_{{\rm eff}}}\left(\frac{t_0-t}{5GM_c/c^3}\right)^{-1/4} \cos\left[2\phi_0+2\phi(t-t_0;M_c,\eta)\right],
\end{equation}
where $t_0=t_c+\tau(\vec{x}(t))$ is the coalescence time at the detector,
$\phi_0$ is given by
\begin{equation}
 2\phi_0=2\phi_c-{\rm arctan}\left(\frac{F_{\times}(\theta,\phi,\iota,\psi;t)}{F_{+}(\theta,\phi,\iota,\psi;t)}\frac{2{\rm cos}\iota}{1+{\rm cos}^2 \iota}\right),
\end{equation}
and the effective luminosity distance to the source, $D_{\rm eff}$, is given by
\begin{equation}
D_{{\rm eff}}=d_L \left[F^{2}_{+}\left(\frac{1+{\rm cos}^2 \iota}{2}\right)^2+F^{2}_{\times} {\rm cos}^2 \iota
\right]^{-1/2}.
\end{equation}

The fourier transformation of the strain~\eqref{eq:strain} can be obtained using the stationary phase approximation.
For a ground-based GW detector, $F_{+}$, $F_{\times}$, and $\tau$ in~\eqref{eq:strain} can be regarded as constants
for a GW burst.
In this case, the frequency-domain version of the strain reads
\begin{equation}
\label{eq:strainf}
\tilde{h}(f)=-\left(\frac{5\pi}{24}\right)^{1/2}\left(\frac{GM_c}{c^3}\right)\left(\frac{GM_c}{c^2 D_{{\rm eff}}}\right)\left(\frac{GM_c}{c^3}\pi f\right)^{-7/6}e^{-i\Psi(f;M_c,\eta)},
\end{equation}
where $\Psi$ is written to the second PN order by
\be
\Psi(f;M_c,\eta) &=& 2\pi ft_0-2\phi_0-\frac{\pi}{4}+\frac{3}{128\eta}\left[\nu^{-5}+\left(\frac{3715}{756}+\frac{55}{9}\eta\right)\nu^{-3}\right. \nonumber \\
&& \left.-16\pi\nu^{-2}+\left(\frac{15293365}{508032}+\frac{27145}{504}\eta+\frac{3085}{72}{\eta}^2\right)\nu^{-1}\right], \\
\nu &=& \left(\frac{G\pi M}{c^3}f\right)^{1/3}.
\ee

For space-based GW detectors such as LISA and Taiji,
the observation time for a GW signal lasts for several days, months or even years.
Thus, the location change of the source to the detector cannot be ignored.
In general, $F_{+}(t)$, $F_{\times}(t)$ and $\tau(\vec{x}(t))$ in~\eqref{eq:strain} are functions of observation time.
According to the forward modeling of LISA described in Ref.~\cite{Rubbo:2003ap},
to linear order in eccentricity, the time delay is given by
\begin{equation}
\tau(t)=-\frac{R}{c}{\rm sin}\theta\,{\rm cos}(\alpha -\phi)-\frac{1}{2}e\frac{R}{c}{\rm sin}\theta\big[\cos(2\alpha-\phi-\beta)-3\cos(\phi-\beta)\big],
\end{equation}
where $R = 1 \, {\rm AU}$, $e$ is the eccentricity of the detector's orbit,
$\beta = 2\pi n/3$ ($n=0,1,2$) is the relative phase of three spacecrafts,
and $\alpha=2\pi f_m t+\kappa$ is the orbital phase of the guiding center.
Like LISA, Taiji is viewed as a combination of two independent detectors in our analysis.
Here $\kappa$ is the initial ecliptic longitude of the guiding center and $f_m=1/{\rm yr}$.
The detector response functions can be written as
\be
F_{+}(t) &=& \frac{1}{2}\Big({\rm cos}(2\psi)D_{+}(t)-{\rm sin}(2\psi)D_{\times}(t)\Big), \\
F_{\times}(t) &=& \frac{1}{2}\Big({\rm sin}(2\psi)D_{+}(t)+{\rm cos}(2\psi)D_{\times}(t)\Big).
\ee
Using the low frequency approximation one has
\be
D_{+}(t) &=& \frac{\sqrt{3}}{64}\bigg[-36{\rm sin}^2\theta\,{\rm sin}\big(2\alpha(t)-2\beta\big) \nonumber \\
&& \left.+\big(3+{\rm cos(2\theta)}\big)\bigg({\rm cos}(2\phi)\Big(9\sin(2\beta)-{\rm sin}\big(4\alpha(t)-2\beta\big)\Big)\right.\nonumber \\
&&\left.+{\rm sin}(2\phi)\Big({\rm cos}\big(4\alpha(t)-2\beta\big)-9\cos(2\beta)\Big)\bigg)\right.\nonumber \\
&& -4\sqrt{3}{\rm sin}(2\theta)\Big({\rm sin}\big(3\alpha(t)-2\beta-\phi\big)-3{\rm sin}\big(\alpha(t)-2\beta+\phi\big)\Big)\bigg],\\
D_{\times}(t) &=& \frac{1}{16}\bigg[\sqrt{3}{\rm cos}\theta\Big(9{\rm cos}(2\phi-2\beta)-{\rm cos}\big(4\alpha(t)-2\beta-2\phi\big)\Big)\nonumber\\
&& -6{\rm sin}\theta\Big({\rm cos}\big(3\alpha(t)-2\beta-\phi\big)+3{\rm cos}\big(\alpha(t)-2\beta+\phi\big)\Big)\bigg].
\ee
The stationary phase approximation is employed to obtain
the frequency-domain version of the strain given by the same as~\eqref{eq:strainf},
in which $F_{+}$, $F_{\times}$ and $\tau$ are replaced by~\cite{Zhao:2017cbb}
\begin{equation}
F_{+}(f)=F_{+}(t=t_f), \quad F_{\times}(f)=F_{\times}(t=t_f), \quad \tau(f)=\tau(t=t_f),
\end{equation}
where
\begin{equation}
\label{eq:tf}
t_f=t_c-\frac{5GM}{256\eta c^3}\left[\nu^{-8}+\left(\frac{743}{252}+\frac{11}{3}\eta\right)\nu^{-6}-\frac{32\pi}{5}\nu^{-5}
+\left(\frac{3058673}{508032}+\frac{5429}{504}\eta+\frac{617}{72}{\eta}^2\right)\nu^{-4}\right].
\end{equation}
In our analysis, we only consider the leading term in Eq.~\eqref{eq:tf}.
Like LISA, Taiji consists of a triangle of three identical spacecrafts in the heliocentric orbit.
Therefore, these results apply to Taiji.

\noindent{\bf Fisher information matrix approach.}
If the strain is well modeled by the formulas obtained above,
the parameter estimation from maximum likelihood test is close to the true value of the parameters
and the errors can be estimated by the Fisher information matrix. For a network including $N$ independent detectors, the Fisher information matrix can be written as
\begin{equation}
\label{eq:fisherm}
\Gamma_{ij}=\left(\frac{\partial_i \boldsymbol{d}(f)}{\partial \lambda_i}, \frac{\partial_j \boldsymbol{d}(f)}{\partial \lambda_j}\right),
\end{equation}
where $\boldsymbol{d}$ is given by
\begin{equation}
\boldsymbol{d}(f)=\left[\frac{\tilde{h}_1 (f)}{\sqrt{S_1 (f)}},\frac{\tilde{h}_2 (f)}{\sqrt{S_2 (f)}},\cdots,\frac{\tilde{h}_N (f)}{\sqrt{S_N (f)}}\right]^{\rm T},
\end{equation}
and $\lambda_i$ denote the parameters ($M_c$, $\eta$, $d_L$, $\theta$, $\phi$, $\iota$, $t_c$, $\phi_c$, $\psi$) for nonspinning MBHBs.
Here, $S_i (f)$ is the noise power spectral density of the $i{\rm th}$ detector and $\tilde{h}_i (f)$ is the strain on it.
The bracket in~\eqref{eq:fisherm} for two functions $a(t)$ and $b(t)$ is defined as
\begin{equation}
\label{eq:filter}
(a,b)=2\int_{f_{\rm low}}^{f_{\rm up}}\left\{\tilde{a}(f)\tilde{b}^{*}(f)
+\tilde{a}^{*}(f)\tilde{b}(f)\right\}{\rm d}f
\end{equation}
In our analysis, we choose $f_{\rm up}$ as the innermost stable circular orbit (ISCO) frequency $f_{\rm isco}$, which is given by
\begin{equation}
f_{\rm isco}=\frac{c^3}{6\sqrt{6}\pi GM}.
\end{equation}

The root mean square errors can be estimated by the Fisher information matrix
\begin{equation}
\sqrt{\langle \Delta \lambda_i^2 \rangle} =\sqrt{(\Gamma^{-1})_{ii}}.
\end{equation}
Since there are nine parameters for a nonspinning MBHB, the Fisher information matrix is a $9\times9$ matrix.
The sky location of the GW source is described by the sky coordinates $(\theta,\phi)$ and the luminosity distance $d_L$.
The error in solid angle is given by
\begin{equation}
\Delta \Omega_s=2\pi |{\rm sin}\theta | \sqrt{\langle \Delta \theta^2 \rangle \langle \Delta \phi^2 \rangle-\langle \Delta \theta \Delta \phi \rangle^2},
\end{equation}
where $\langle \Delta \theta^2 \rangle$, $\langle \Delta \phi^2 \rangle$ and $\langle \Delta \theta \Delta \phi \rangle$ are given by the inverse of the Fisher information matrix.
In our analysis, we focus on the angular resolution and luminosity distance uncertainty.
Although the Fisher information matrix gives a lower limit for parameter estimation,
it is very helpful to estimate the localization capability for future experiments.

\noindent{\bf Mock data generation.}
We generate the mock data
assuming a flat $\Lambda$CDM cosmology with $\Omega_m=0.31$, $\Omega_\Lambda=0.69$, and $H_0=67.74$ km s$^{-1}$ Mpc$^{-1}$~\cite{Ade:2015xua}.
Given a source redshift,
we can easily calculate the luminosity distance and angular diameter distance to the source.
Without loss of generality, we consider equal-mass black hole binaries
with the total intrinsic masses of $10^5 \, M_\odot$ and $10^6 \, M_\odot$.
Note that the observed mass $M_{\rm obs}$ is related to the intrinsic mass $M_{\rm int}$ by the relation $M_{\rm obs}=(1+z)M_{\rm int}$.
Since the intrinsic mass is degenerate with the redshift from GW measurements,
the observed mass is used in our analysis.

LISA consists of a triangle of three spacecrafts separated by 2.5 million kilometers
while Taiji has three spacecrafts with 3 million kilometers separations.
Compared to LISA, Taiji is more sensitive to low-frequency GWs.
In our analysis, we adopt the noise power spectral density for LISA obtained in~\cite{Audley:2017drz}
and for Taiji described in~\cite{Guo:2018npi}.
The coalescence of MBHBs in general lasts for several days, months or even years in the frequency band of LISA and Taiji.
With the noise power spectral density, we calculate~\eqref{eq:filter}
choosing $f_{\rm low}=0.4$ mHz for the binary with a total intrinsic mass of $10^5 \, M_{\odot}$, $f_{\rm low}=0.1$ mHz for the binary with a total intrinsic mass of $10^6 \, M_{\odot}$ and $f_{\rm low}=0.03$ mHz for the binary with a total intrinsic mass of $10^7 \, M_{\odot}$.

The detector response functions and time delay between LISA and Taiji
depends on the relative position of two detectors via $\alpha=2\pi f_m t+\kappa$,
which indicates that the angular resolution varies with the configuration angle $\beta$.
Given a redshift and total intrinsic mass of equal-mass black hole binaries,
the GW signals are generated with random binary orientations and sky directions.
To investigate the effect of the configuration angle on the angular resolution,
we simulate 10,000 random MBHB samples with the total intrinsic mass of $10^5 \, M_{\odot}$ and redshifts of $z=1$ or $z=3$,
assuming that $\kappa=0$ for LISA and $\kappa$ for Taiji is chosen in the range of $[0,\pi]$.
Moreover, the sky location, binary inclination, polarization angle and coalescence phase
are randomly chosen in the range of $\theta \in [0,\pi]$, $\phi \in [0,2\pi]$, $\iota \in [0,\pi]$, $\psi \in [0,2\pi]$ and $\phi_c \in [0,2\pi]$, respectively.
Without loss generality, the coalescence time $t_c$ is set to be zero in our analysis.


\end{document}